# Babylonian-style Programming

## Design and Implementation of a General-purpose Editor Integrating Live Examples Into Source Code

David Rauch[a], Patrick Rein[a], Stefan Ramson[a], Jens Lincke[a], and Robert Hirschfeld[a]

a    Hasso Plattner Institute, University of Potsdam, Germany

**Abstract**    When working on a program, developers traditionally have to simulate the behavior of the abstract code in their heads until they can execute the application. Live programming aims to support the development and comprehension of programs by providing more immediate feedback on program behavior, but the divide between code and behavior often remains. The goal of example-based live programming is to remove this gap by allowing programmers to explore the actual behavior of their code during development. This is achieved by defining live examples for parts of the application.

The idea of live examples has been already addressed in other tools and environments. However, most of those solutions are limited to specific domains and are suitable only for small programs. Thus, we aim to extend the application of example-based live programming to more complex programs potentially spanning multiple modules.

We investigate existing solutions to derive a set of requirements for an integration of live examples into source code. Based on these requirements we propose a new approach to live examples and present a prototype in its support. We reproduce, discuss, and extend scenarios from related work to show the practicality of our approach in the context of larger, more complicated, and with that also more realistic scenarios. Also, we measure and evaluate the system response time of our prototypical implementation.

Our first results show that example-based live programming can provide more insights into the run-time behavior of parameterized code for non-trivial programs. They also reveal unsolved and new challenges affecting example-based live programming environments.

In presenting this more general approach to example-based live programming, we hope to motivate further research into this area and to make practical solutions available.

## The Art, Science, and Engineering of Programming

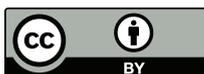

## 1 Introduction

A fundamental cognitive challenge in programming is the gap between a program's code and its behavior [20]. When developing a program, developers may consider *concrete* data such as exemplary input and expected output of a function in order to formulate the *abstract* program. While the abstract nature of the program allows it to go beyond the concrete data considered by its developers, this process of abstraction also means that the program is no longer tightly connected to its real-world use-case. To quote Edwards [4]:

> Software is abstract: this is the source of both its power and its problems.

Live programming systems aim to support the development and comprehension of programs by providing more direct and immediate feedback about the program's behavior to developers [30]. This feedback is often realized by presenting developers with some of the intermediate results and effects of the program under development. In some situations, developers may even be able to inspect certain parts of the program or see intermediate results. Cognitive science also suggests that abstract concepts are typically understood through metaphor and more concrete concepts [18].

This approach is particularly useful when developing a program with the sole purpose of producing a single specific result, such as an image. However, many applications are not developed to obtain just one result. While developing a program for the sole purpose of drawing one specific image is a realistic use-case, one may also imagine developing a program that acts as an abstract framework for drawing arbitrary images depending on its parameters, or one that processes data from an external source.

This dependency on additional input to fill parameters represents a challenge for live programming systems, as it is no longer possible to simply execute the program and see the result. Instead, developers now have to provide example invocations for the parameterized program in order to examine its behavior. Further, programmers may only work on a small, isolated part of the program and may therefore not be interested in the behavior of the entire application.

Live programming with explicitly defined examples aims to reconnect abstract programs with their concrete, real-world use-cases by allowing programmers to express their, so-far imagined, examples as a part of the application. The environment can make use of these examples to continuously provide live feedback, for example on the resulting runtime state. By doing so, developers can see how the program behaves in concrete use-cases without having to explicitly execute it or imagine the process in their heads [4], as visible in figure 1. This allows for quick, relevant, and concrete feedback during programming.

Explicit examples could therefore enable valuable mechanisms to support the development, comprehension, and exploration of applications. Complex applications with many layers of abstraction could especially benefit from explicit examples, as it can add concrete information to abstract code.

However, complex applications may present a challenge for live programming with explicit examples, as individual components may be dependent on each other or assume to be executed in certain environments. Many existing systems with explicit examples are therefore limited in scale, often by focusing on small applications contained in one module [33, 13], for example within an educational context [36, 15], or by only supporting specific domains, such as applications that generate images using a specific framework [36, 35].

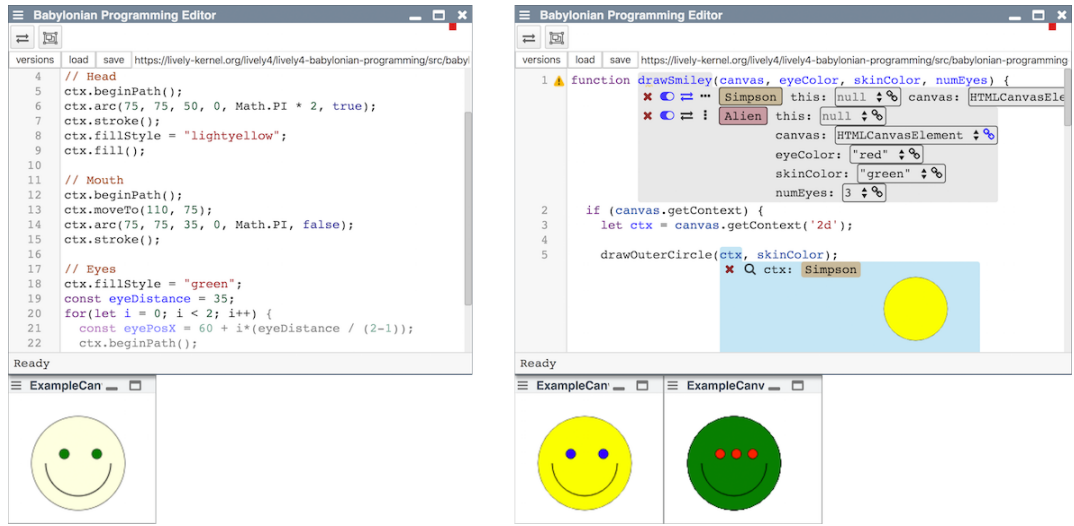

**Figure 1** Live results for a concrete implementation (left) and an abstract implementation with live examples (right)

We therefore examine existing solutions and develop a new general-purpose tool design for the integration of examples into applications with complex code bases.

**Contributions** This work contributes a survey of a sample of existing Live Programming Environments (LPEs) with explicit and implicit examples, with a focus on their applicability to applications with extensive code bases. We examine how different environments leverage live examples to support the development, comprehension, and exploration of programs. Through this examination, we identify common requirements for LPEs with explicit examples supporting programmers while working with complex applications.

We then use the results of this survey to create a new design for the integration of live examples into object-oriented source code which scales to more extensive code bases spanning several modules[1]. An implementation of this design is provided. We further report on the programming experience with our implemented tool when working in prototypical scenarios from related work. We also provide a preliminary evaluation of the resulting response time of the tool and compare its features to the initial requirements and the features of other environments.

**About the name** The potential advantage of concrete examples appears to be old knowledge. Some of the earliest surviving documents dealing with computation, nearly

---

[1] See https://doi.org/10.5281/zenodo.2541913 for a demonstration of the resulting tool.

**Babylonian-style Programming**

4000 years old stone tablets from the Babylonian period (approx. 1800 BCE–500 BCE), already show descriptions of algorithms with integrated examples [16].

**Outline**   The remaining paper is structured as follows. We begin by evaluating and comparing existing environments in section 2. We then present a new design targeting complex applications in section 3 and describe its implementation in section 4. The presented design is evaluated in section 5 before summarizing our findings and discussing possible directions for future work in section 6.

## 2   State of the Art

There are a number of systems aiming to bring live feedback to programming. For the context of this work, we will distinguish between LPEs with *implicit* and *explicit* examples. Systems with implicit examples require programmers to modify the application itself in order to define examples. Using explicit examples, programmers can keep source code abstract while also providing concrete example invocations. These examples may be specified using a special syntax or User Interface (UI), and are used to obtain concrete information for abstract source code.

Generally, examples are possible invocations of functions. As programmers modify the source code, these functions are automatically invoked with the specified parameters in order to obtain updated results. As examples usually do not retain state across invocations, they should provide all information necessary to invoke their function in the required context. Examples are therefore conceptually similar to unit tests [2, 4, 32]. However, unit tests aim to test code for conformance, whereas live programming with examples focuses on visualization techniques to make the described dynamic behavior more concrete [4].

Many existing solutions are focused on a specific domain, which often limits their applicability to general-purpose programming. We will therefore examine the applicability of existing works to complex applications and summarize identified requirements.

### 2.1  Live Programming Environments with Examples

**Example Centric Programming**   One early environment with implicit examples [4] presents a two-pane design, with the right pane containing a standard code editor in which both code and examples can be edited, and the left pane containing a trace of the execution of the examples (see figure 2). Examples are defined as standalone pieces of code that call the functions under consideration, and they are generally appended at the end of the file.

As examples are simply defined at the end of the file, there is no syntactic or visual difference between normal code and examples. This may cause problems when code is loaded in a system which is not aware of the examples. The location of the examples also physically distances them from the functions they refer to, forcing programmers to jump between methods and their examples, especially in larger files. The same is

David Rauch, Patrick Rein, Stefan Ramson, Jens Lincke, and Robert Hirschfeld

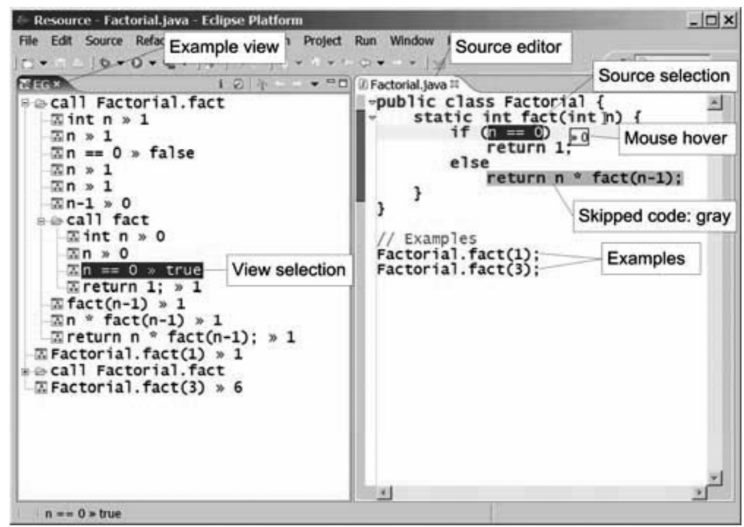

**Figure 2** The two-pane layout of Example Centric Programming [4]

```
var low = 0;
var high = 5;

while (low <= high) {
   p(low, [0,1,2,3,4,5]);
   p(high, [5,5,5,5,5,5]);

   low++;
}
```

**Figure 3** Probes in Live Literals which are actual calls to the function p [33]. The probe displays the value of the expression for the first argument by changing the literal displayed as the second argument. The literal presumably passed as the second argument is thus not an actual argument but a literal whose value depends on an example execution of the probe.

also true for behavioral information: As it is only shown in the left pane, there is a gap between the code's syntax and its behavior.

Some information is, however, displayed directly in the editor. Selecting an element in the execution trace will also select the corresponding code in the editor, and code that was not executed by an example is faded out.

**Live Literals** A more recent example for explicit examples aims to make source code more live and interactive by introducing *live literals* [33], which are part of the source code itself (see figure 3 and figure 4). Developers can call special functions such as run within functions to define exemplary invocations. Once such an example has been defined, the function p can be called to create a *probe*. The expression passed to the function p will be evaluated, and a live result will be presented. This allows

**Babylonian-style Programming**

```
function factors(n) {
  run({n: 2});
  test([
    {n: 0, result: []},
    {n: 4, should: [1,4,2], result: true}
  ]);
  var fs = [];
  for (var i = 1; i <= Math.floor(Math.sqrt(n)); i++) {
    if (n % i === 0) {
      fs.push(i);
      if (n / i !== i) fs.push(n / i);
    }
  }
  p(fs, [1,2]);
  return fs;
}
```

■ **Figure 4** The *test* and *run* functionality of Live Literals [33]. The first fields in the object literal of the argument to test denote the arguments passed to the function and the should field can contain an expected value. If a should field is present, the result field indicates whether the function returned the expected value, if the should field is not present, the result field shows the result value of the function. The *run* call provides the example used to evaluate the probes, such as the one at the end of the function.

programmers to inspect the state of their application for a given example at a given point.

The visual presentation of probes is limited by the fact that they are inserted as text into the source code. For example, if an expression is executed multiple times, the probe shows all values in an array. This is ambiguous as the same visual representation is also given when an expression is executed only once but results in an array. Additionally, the design only features probes which support one example at a time, which means that old examples have to be regularly removed. It is possible to create multiple test calls, but these behave more like unit tests and only show the return value of the function.

**Shiranui** Another system with explicit examples focuses on combining the quick feedback of live programming with the persistence of unit testing [13]. Shiranui is both a dynamically typed, functional programming language, and a development environment (see figure 5). It is based on a workflow in which programmers define examples while developing a new function. The resulting examples can be persisted as unit tests afterwards.

Examples are defined as text and they describe exemplary invocations of functions with expected return values. Programmers can *focus* on a specific example, which will cause code relevant to the example to be printed in bold text. It is also possible to define probes, which behave similar to the probes in Live Literals [33]. One differentiating feature of probes in Shiranui is that it is also possible to focus on individual results of expressions. After focusing on one result a dynamic slice for that result is displayed [1, 17]: all intermediate results that led to the result are printed on the right-hand-side and unexecuted code is struck through (see line 7 in figure 5 for a selected result).

```
1 #+ fib(1) -> 1;                    n at [121,122] = 4
2 #- fib(3) -> 3;                    n at [130,131] = 4
3 #- fib(4) -> 5 || 4;               fib at [167,170] = <|a=$(fib->a)fib|>
4                                    n at [171,172] = 4
5 // NOTE: 1 1 2 3 5                 fib(n-1)  at [167,176] = 3
6 let fib = \fib(n){
7     #* n -> 4,3,2,1;
8     if n = 0 or n = 1{
9         1;
10    }else{
11        fib(n-1) + 1; //BUG!
12    }
13 };
```

■ **Figure 5** A screenshot of the Shiranui environment [13]. At the top three example calls to the function fib are given. The third one is currently focused on and returns the value 4. The value 5 is the expected value provided by the programmer of the example. In line 7 a probe is shown on n. In the list of values (shown after "->") the value 4 is selected (illustrated by the underlining) and thus the editor strikes out any code sections which are not part of the trace for the call with this argument.

```
function binarySearch (key, array) {      key = 'g'
    var low = 0;                          array = ['a','b','c','d','e','f']
    var high = array.length − 1;          low = 0
                                          high = 5
    while (low <= high) {
                                          low  = 0 |  3 |  5
        var mid = floor((low + high)/2);  high = 5 |  5 |  5
        var value = array[mid];           mid  = 2 |  4 |  5
                                          value = 'c' | 'e' | 'f'
        if (value < key) {
            low = mid + 1;                low  = 3 |  5 |  6
        }
        else if (value > key) {
            high = mid − 1;
        }
        else {
            return mid;
        }
    }
    return −1;                            return −1
}
```

■ **Figure 6** An implementation of binary search in Inventing on Principle [35]. The left part shows the source code while the right side shows values from the execution of the example arguments given in the first two lines. The values of the while loop are organized in one column per iteration.

This may help developers to identify not just that a result is incorrect, but also why it is incorrect.

**Inventing on Principle: General Code Demonstration**   In Inventing on Principle [35], Victor presents several different designs aimed at making programming more interactive and more immediate. While most designs in this talk are aimed at specific domains, one design has a more general-purpose approach based on explicit examples (see figure 6).

Programmers are presented with a source code editor on the left, and an additional panel on the right that displays intermediate values. As this design has no support for probes, the right-hand-side panel shows all values as they are assigned. If multiple values exist for the same statement, they are displayed in columns. While this helps

**Babylonian-style Programming**

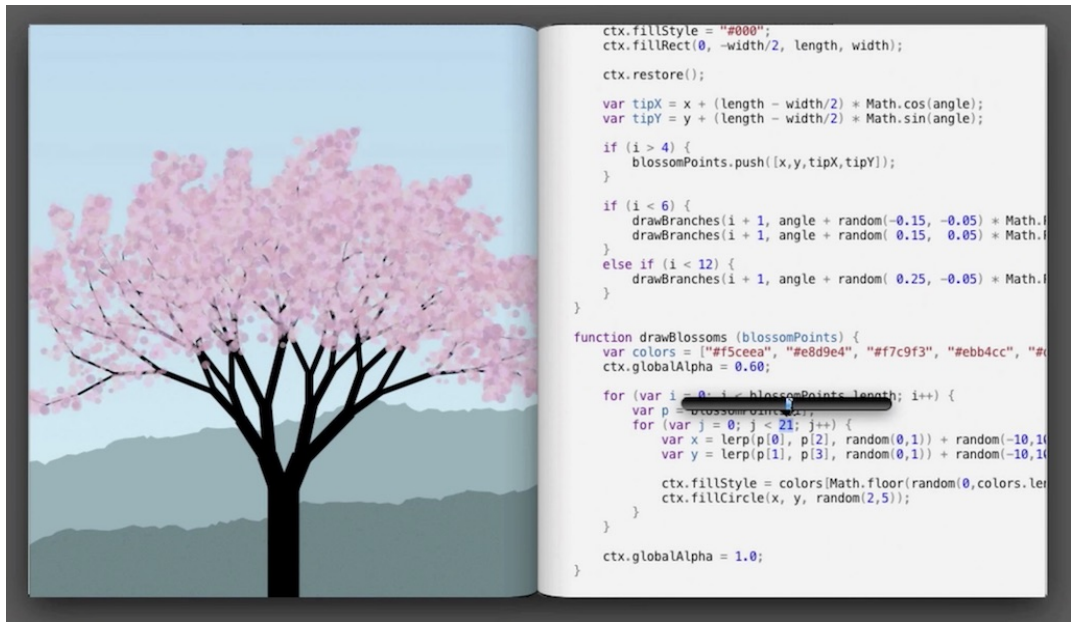

**Figure 7** The canvas demo from Inventing on Principle [35]. The picture on the left is generated by the program on the right. In the screenshot the user is manipulating a value using a slider. While dragging the slider the picture on the left is continuously updated to represent the result from the program execution including the changed value.

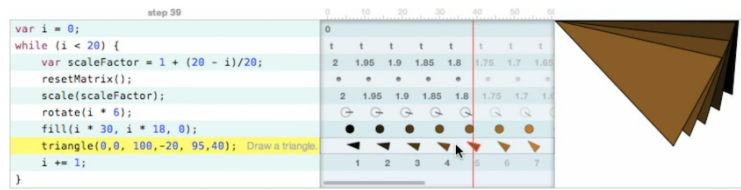

**Figure 8** Inspecting the state of a program at a given point in time using Learnable Programming [36]

to identify values of the same iteration, it does not scale to navigating and exploring larger examples with hundreds of iterations. Support for complex objects is also not presented, limiting this design to small applications.

**Inventing on Principle: Canvas Demonstration**  One of the designs based on implicit examples presented in Inventing on Principle [35] shows a specialized editor with an integrated drawing area (see figure 7). The right side shows the source code for drawing a concrete scene, while the left side shows the current image drawn by the code. Programmers can modify numerical values by dragging them and they can select areas in the image to jump to the code that produced them. This design does not offer a way to show intermediate values. As the focus is on presenting graphical output, there is also no support for other kinds of objects.

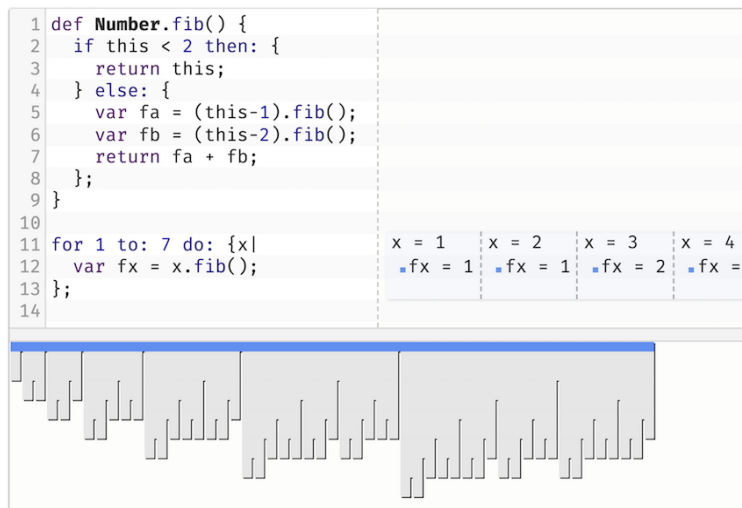

■ **Figure 9** The three-pane layout of Seymour [15]. The right side of the tool shows the values during the execution which is triggered by the example calls at the bottom of the editor on the left side. The macro visualization at the bottom allows users to navigate through the activations of the fib function.

**Learnable Programming** In Learnable Programming [36], Victor presents a design for a source code editor aimed at teaching the basics of programming through small, drawing-based tasks with one implicit example (see figure 8). Because of its limited domain, the design does not support explicit probes, but instead shows all values over time. This is useful for the small examples presented, but may not scale to situations in which, for example, statements in a loop are executed thousands of times. The design is also tightly integrated into the specific language and framework, as it offers specialized visualizations for the supported geometric objects, but does not support arbitrary objects.

While it still follows the general idea of the two-pane layout, the design goes beyond by integrating several aspects directly into the source code editor. For example, the ability to scrub though iterations of a loop directly in the editor makes program flow significantly more tangible.

**Seymour: Live Programming for the Classroom** Seymour [15] is a live programming system focused on teaching programming in a classroom setting (see figure 9). At its core, Seymour's design has three components: A code editor, a *micro* visualization which provides details about the control flow and different values across function activations, and a *macro* visualization which provides an overview of the overall program execution. Example invocations are typically added at the bottom of the program.

In the micro visualization, Seymour presents state changes, both through assignments and side-effects. The table-like layout, in which every iteration step occupies one column, allows programmers to correlate values within iteration steps and supports reasoning such as "when $x$ was 3, $f_x$ was 2". As Seymour does not support explicit

```
    render();
    stats.update();
}
var counter = 0;
var direction = 1;
function render() {
    var time = Date.now() * 0.001; 1377199075137

    mesh.rotation.x = time * 0.25; { _x: 344299768.78025,
                                     _y: 688599537.5605,
                                     _z: 0,
                                     _order: 'XYZ',
                                     _quaternion: [Object] }
    mesh.rotation.y = time * 0.5; 688599537.5685

    renderer.render( scene, camera );
```

■ **Figure 10** Three different watches in the Light Table editor [9]. All three show the result of the underlined expression next to them. The first and the last watch show a number while the second one shows an object.

probes but probes all values, the design may not scale to more complex programs, as it becomes difficult to focus on specific values.

One aspect in which Seymour goes beyond other designs is that it considers the relevance of both *object value* and *object identity* to developers, while most other systems only focus on presenting object values. Object identity is displayed as an emoji character alongside object values.

**Light Table**   Light Table is an Integrated Development Environment (IDE) focused on providing instant feedback, visualizing data flow through code, and being highly customizable [9]. It differs from most other examined designs as it is a general-purpose editor rather than a design focused on a specific task (see figure 10). One of the key features of its design is the concept of watches, which are similar to probes. In this regard they are similar to watch expressions offered by debuggers in some development environments. Watches in Light Table are, however, not separate statements but UI elements attached directly to syntax elements of the program. Once attached, they display the most recent value of their expression. Again Light Table requires programmers to manually start the execution of their program through an implicit example.

The design also supports the concept of live documents. In this mode, programmers can write code in one panel and see the full source code annotated with concrete, traced values in the secondary panel. This behavior is also supported across files, meaning that programmers can open additional panels for functions called from their code, which will also be displayed with live values. The documents presented in the secondary panel can therefore be considered a concrete version of the abstract code written in the editor.

However, as there is no support for examples, all live results come from simply executing the entire document. This limits live documents to small programs or abstract modules annotated with example calls.

David Rauch, Patrick Rein, Stefan Ramson, Jens Lincke, and Robert Hirschfeld

## 2.2 Conclusions

Comparing different live programming systems, both with implicit and explicit examples, we find that many systems are designed for specific domains. Especially the investigated designs that aim to tightly integrate live feedback for a specific domain, such as rendering images, into the code editor are often, in the end, limited to that domain. At the same time, we argue that the idea of live programming with examples can also be applied to complex applications. Comparing the existing solutions and their limitations, we can identify a number of requirements to support the development of such applications:

**Feedback on Runtime State**  In order to explore and understand the effects of changes to the program, programmers should be able to see and inspect all parts of the runtime state, including the results of single expressions [20]. This in turn entails the following requirements:

- **Feedback granularity** When exploring behavior for a particular task, programmers might only be interested in the results of selected expressions and not in the complete runtime state of one part of the application [4, 9, 13, 33]. Further, they might even be interested in feedback on additional expressions which are not part of the ordinary program code [33].
- **State over time** As larger applications tend to have complex control flow structures, the exact evolution of runtime state is not always immediately clear from reading the source code. As live programming aims to bridge the gap between our mental model and the actual behavior, the transition of runtime state should become visible [4, 15, 35, 36].
- **State over modules** To fully understand the behavior of an application, programmers often investigate not only one part of the application. For example, a programmer exploring the behavior of a method might also look into the runtime state of other methods called by the currently investigated method. Programmers should be able to see the effects of one part of the application in other parts of the application even if they reside in other modules.
- **Arbitrary objects** As most real-world applications use complex objects and large data structures, a scalable design has to support developers by providing relevant, yet concise information about these objects [9, 15].
- **Domain-specific feedback** While applications can involve arbitrary objects, programmers might work on parts of an application which has a particular domain. If there is a visualization of these domain objects which could be useful to the programmer, a programming environment should provide it or let the programmer configure it [36].

**Associating Examples with Code**  As the goal of using examples it to ease the process of getting feedback on specific parts of an application, programmers should be able to associate examples with parts of an application in various ways.

**Babylonian-style Programming**

**Multiple examples for one part of the application**  In order to cover a variety of different traces of one part of an application, programmers might want to use different examples simultaneously, such as an example for the ordinary case and one for an erroneous case [4, 13, 33].

**Reusing parts of examples**  Some part of an example might cover a number of central aspects of an application, for example an example instance of a class might be relevant for the ordinary case of most methods of a class. The programming environment should allow programmers to reference such example instances in different examples throughout an application to save programmers the mental effort of regularly having to learn the concrete data of a new example, similar to fixtures in unit tests [2, 25].

**Determining Relevant Sections of Code**  When understanding a complex part of an application, programmers might only be interested in the lines of the code relevant to the current task. The programming environment should support programmers with spotting the relevant lines or navigating to these lines. The relevance of a line depends on the question at hand:

**Control flow**  Given that the actual control flow of a program might not always be obvious, programmers might ask which lines were actually executed [4].

**Runtime state**  When programmers try to understand a particular runtime state they might want to know which statements lead to a particular runtime state [13].

**Program output**  In some instances, programmers might be interested in the side-effects of the execution of a given section of code, for example graphical output. In that case, programmers might ask which statements lead to a particular program output [36].

**Specifying Context**  When working in complex systems, code might depend on the presence of other structures, such as special objects, for example caches, or external resources, such as databases. Code depending on these structures can only be understood if these structures are present. This is similar to the challenges automated tests face, for example when dealing with objects the unit under test depends on [2, 25]. To limit the scope of elements to be understood by the programmers, the programming environment should allow them to reuse execution contexts or mock resources. To limit the scope even further, the programming environment should execute examples isolated from each other similar to the way automated tests should be executed.

**Keeping Track of Assumptions**  One part of a programmers mental model of parts of an application might be assumptions about invariants that should hold throughout the execution. For example, when writing unit tests these assumptions are documented through assertions [2, 25]. Programming systems often provide some means to add assertions into application code. Additionally, an environment supporting explicit examples should provide programmers with means to specify assertions specific to examples. These assertions should cover individual expressions, as well as the overall execution of a part of the application [13, 33]. Further, the environment should provide feedback to the programmers about any erroneous execution.

**Navigating the Trace** An example invocation might create a complex trace covering many parts of an application spread across several modules [15, 24]. The trace might include ordinary method or function calls but also loops or recursive calls. A complex trace can make it difficult for programmers to understand, for example, where a method call came from or which modules an example will cover. This becomes particularly interesting when several different paths trough the trace can lead to the same method and programmers want to investigate how the method is used in one particular case in the trace of the example.

**UI Design Challenges** Besides these requirements, we can also identify two UI challenges for integrating examples into a programming environment:

- Should the system display all information in one *single pane*, or split source code and examples into *two different panes*? Many of the designs presented in section 2 choose a dual-pane layout. This has the advantage that the code itself remains compact. It does, however, also have the disadvantage of visually separating behavioral information from syntax, therefore decontextualizing it [24, 20].
- If a single-pane layout is chosen, should example-related information be *integrated into the source code* as actual statements (see Live Literals), or should it be *displayed on top by the IDE* (see Light Table)? Both require IDE support to provide a live programming experience. The integration into source code is more portable, as examples are literally part of the document. It does, however, limit the visual representation of behavioral information to a text-based format. IDE integration, on the other hand, allows probes and examples to be UI widgets, while requiring special serialization to persist examples. It also potentially allows the same design to be applied to different languages.

## 3 Design

Having examined existing solutions, figure 11 presents a new design for the integration of live examples into source code. The presented design targets complex applications spanning several modules and is guided by the core requirements identified in section 2.2.

The UI is designed around the principle that the majority of additional information is presented in the form of UI widgets that are attached to syntactic elements of the source code. These widgets are called *annotations*, and they follow and display information related to the elements they are attached to. As the design is a single-pane layout, widgets are directly embedded into the source code editor. There are only two permanent buttons in the top toolbar: one to configure the pre- and postscript for the current module (see section 3.5) and one to configure custom instances (see section 3.3). The remaining buttons in the top toolbar are context sensitive: Depending

**Babylonian-style Programming**

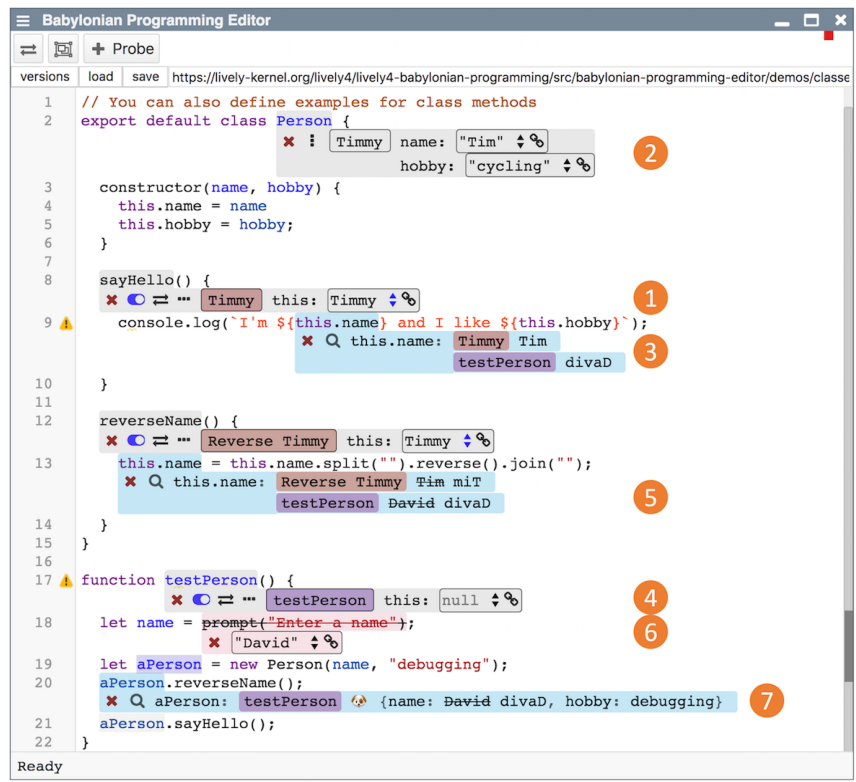

■ **Figure 11** The editor implementing our tool design, showing a local example (1) using a named instance template (2), probes (3, 5) showing results from two different examples (1, 4), a replacement (6), and a probe showing changes on an object (7).

on the current selection, the editor provides buttons to add or remove annotations that are applicable to the selected syntax element.

The following sections illustrates how the features identified in section 2.2 are adopted and adapted in the presented design to match some of the identified requirements.

### 3.1 General UI Design Decisions

There are a number of fundamental differences in the designs of existing LPEs with explicit examples, largely concerning the way in which examples are defined and explored by programmers.

In order to tightly integrate behavioral information into abstract source code, we have chosen a single-pane layout. The presented design is also integrated into the IDE rather than the language itself, as this allows for a more complex visual representation of information.

As we target complex applications, the evaluation of one example will likely involve code in multiple classes and modules. In the presented design, developers can therefore open multiple files simultaneously in different editors. If an example defined in one editor causes code to be executed that is currently opened in another editor, the

```
person.hobby = "testing";
✘ Q person: ▇ 🔮 {name: David, hobby: debuggging testing}

power *= power;
✘ Q power: Powers of 2  2 4 │ 16 │ 256 │ 65536 │ 4294967296
           Powers of 5  5 25 │ 625 │ 390625 │ 152587890625 │ 2.3283064365386964e+22
```

**(a)** Two probes: One showing the properties of an object, the other showing multiple values for different examples

```
for(let i = 0; i < 5; i++) {
✘ |━━━━━━━━━━━ all 5
  outerSum += i;
  ✘ Q outerSum: 0 │ 1 │ 3 │ 6 │ 10
  for(let j = 0; j < 3; j++) {
  ✘ |━━━━━━━ 6 of 15
    innerSum += j;
    ✘ Q innerSum: 4 6
  }
}
```

```
let celcius = prompt("Temperature to convert:");
              ✘ 24 ⇕ %
let fahrenheit = celcius * 9/5 + 32;
                ✘ Q fahrenheit: 75.2
```

**(b)** Two sliders, one of which has been set to a particular iteration

**(c)** A replacement is used to replace user input with a fixed value for example execution

```
function deepCopy(obj) {
  ✘ ◐ ⇄ ⋯ Plain      this: null ⇕ %  obj: {name: "My name"} ⇕ %
  ✘ ◐ ⇄ ⋯ Recursive  this: null ⇕ %  obj: Identifier Path ⇕ %
  ✘ ◯ ⇄ ⋯ HTML       this: null ⇕ %  obj: <div> ⇕ %
```

**(d)** A function with three named examples, two of which are enabled

```
class TreeScene {
  ✘ ⋮ Cherry at day   stemHeight: 100 ⇕ %
                      stemWidth:  30 ⇕ %
                      treeKind:   TreeScene.CHERRY ⇕ %
                      time:       TreeScene.DAY ⇕ %
  ✘ ⋮ Birch at night  stemHeight: 70 ⇕ %
                      stemWidth:  20 ⇕ %
                      treeKind:   TreeScene.BIRCH ⇕ %
                      time:       TreeScene.NIGHT ⇕ %
```

**(e)** Two instance templates attached to a class, one is named "Cherry at day" and the other "Birch at night".

■ **Figure 12** Different kinds of annotations

secondary editor will display the results of that execution even though it did not initiate the evaluation. This allows developers to define examples for functions at high levels of abstraction and, if necessary, "follow" these examples by seeing the results of expressions in lower levels of abstraction.

### 3.2 Feedback on Runtime State

In order to provide programmers with feedback on the runtime state, the presented design incorporates *probes*. The purpose of probes is to trace values of the selected expression during runtime [24]. Two exemplary probes are shown in figure 12a. The presented design attaches probes to existing syntax elements rather than allowing custom probe expressions, as this ensures that probes are side-effect free. Probes also

**Babylonian-style Programming**

associate values with examples. This means that multiple examples can be enabled simultaneously in order to compare their behavior.

If multiple values are recorded for a probe over time, they are displayed in chronological order, separated by a vertical line. If an identifier's value changes as part of a probed statement, the probe will display both the old and the new value. To provide feedback on arbitrary objects, probes capture the properties of objects and print them in a JavaScript Object Notation (JSON)-like syntax. If the single line per examples is not large enough to display all captured data, programmers can click on a value displayed by a probe to open the selected value in the system's object inspector.

Finally, probes can be placed on any expression in any module in the system. If the expression is evaluated in the course of an example the probe will show the recorded values with the color and name of the corresponding example.

### 3.3 Associating Examples with Code

Examples allow programmers to define *example invocations of functions* and methods by attaching them to those callables and entering the parameters, as visible in figure 12d. Every field also has two additional buttons, one to select an instance template from a dropdown menu, the other to create a link to an external object from the LPE.

Examples are also automatically assigned a color, and programmers may also assign names to examples. Both the color and name of an example is displayed in probes. Examples can be enabled and disabled, may have additional setup and tear down code, and may be shown in a horizontal or vertical layout. This allows for multiple examples to be enabled simultaneously without modifying the code itself.

**Reusing Parts of Examples**   To reuse parts of an example, the design allows programmers to use templates for creating objects or reference existing objects in the system.

For creating example instances of a particular class, programmers can use *instance templates*, as shown in figure 12e. They represent a call to the constructor of the class with concrete values provided for the parameters of the constructor. Programmers can assign a name to such a particular call to the constructor and reference it by that name in examples. There the templates are used to create new instances every time the example is executed. In the editor the templates are attached to the names of classes in the context of a class definition.

In addition to these class-specific instance templates representing calls to the a constructor, developers may also want to use instances that can not be obtained by simply calling a constructor, such as instances created using a factory method pattern [7]. To use such instances, developers can create *custom instance templates* in a separate window by specifying arbitrary instructions in order to create and modify a new instance (see figure 22). These templates can also be referenced in examples and follow the same lifecycle as instances created using the normal instance templates.

On some occasions, developers may need to use instances that persist across multiple evaluations of an example, or that exist outside the application under development, such as a canvas object on which an example drawing function should be executed. As the presented design is integrated into a self-sustaining LPE, this can be achieved by accessing objects that already exist in the system. This access to other objects is realized by so-called *links*. In order to use already existing objects, developers can click on the link icon next to form fields and select an external component. Such linked objects have a different lifecycle, instead of being created for every example execution, they persist across executions.

### 3.4 Determining Relevant Sections of Code

In addition to UI elements, behavioral information also influences how code itself is displayed. As soon as at least one example is enabled, the system traces which code was reached during the execution of enabled examples. If a line of code was not reached for any enabled example, this line of code is faded out. Thus, the editor displays which statements are part of the trace of the example invocation.

This optimizes the UI for programmers' intent, as they can scan large modules and quickly see which functions might be relevant to their current task. It may also be used to localize erroneous control flow, as programmers can quickly see whether or not certain conditionals were executed without having to attach additional probes.

### 3.5 Specifying Context

In some cases, examples may also require additional setup- and tear down routines comparable to those in unit tests [2]. *Pre- and Postscripts* can therefore be used on a per-example basis to specify instructions that should be executed before and after the actual function is invoked. These instructions can access and modify the parameters specified for the example. As the pre- and postscript is not necessary for most examples it is excluded from the usual one-pane layout. Instead, clicking the appropriate button opens a new window in which developers can specify the instructions to be executed, as visible in figure 23.

*Replacements* allow programmers to replace arbitrary parts of the source code during the execution of examples. They are therefore an exception to the general rule that examples should apply to the original application. However, replacements can be useful to avoid computations or resource access which would otherwise be complicated to configure or computationally expensive. Further, they can limit the scope of reasoning for an example. Replacements are similar to mock objects used in unit testing [2].

For example, a function may ask for input from the user as shown in figure 12c. This would require developers to provide the required input every time an example for this function is evaluated. Using a replacement, programmers can replace the expression asking for input with a fixed value. This way, examples can be executed without additional input.

### 3.6 Keeping Track of Assumptions

The presented design does not provide a dedicated way for keeping track of assumptions about the code. It does however indicate whenever an example leads to an

**Babylonian-style Programming**

exception during execution by showing a symbol in the example selection widget. Thus, assert statements can be used to document invariants. As examples are executed separately, a failing assertion in the execution of one example does not affect the others.

### 3.7 Navigating the Trace

In order to navigate iterations and multiple activations of functions, the design supports so-called sliders. Sliders can be attached to flow control structures and functions and allow programmers to *slide* through the iterations or activations. As programmers slide through the trace, probes directly within the selected structure switch from displaying all captured values to displaying only values captured during this iteration or activation. This allows programmers to recreate the execution of the program and to follow data over time. It also makes it easier to reason about the causal connections between values, as only values that were captured during the same iteration are displayed.

## 4 Implementation

The presented design was implemented in JavaScript [3, 6] as a web component in the Lively4 LPE [21] using CodeMirror [10] as a text editor[2]. Lively4 is a web-based self-sustaining development environment with a focus on exploratory-style live programming using web components [21]. As the implementation is executed by a browser without special permissions such as access to the debugger or runtime, the application's source code is transformed in order to integrate probes and examples.

In order to present programmers with live results of their examples, these examples have to be regularly evaluated. As examples are defined for functions and methods, evaluating examples means calling the given function or method with the parameters defined by the example. The overall process comprises the transformation of source code enriched with examples to an executable Abstract Syntax Tree (AST) and the actual execution of examples including tracing and updating the UI.

As there may be multiple editors open at the same time and the execution of examples has to be coordinated between these editors, editors are managed by a central worker, which evaluates enabled examples and afterwards provides all editors with updated results (see figure 13). The parsing and transforming of the AST can be performed on a separate thread, whereas the actual execution has to happen on the main thread as it requires access to the Document Object Model (DOM) [26, 28].

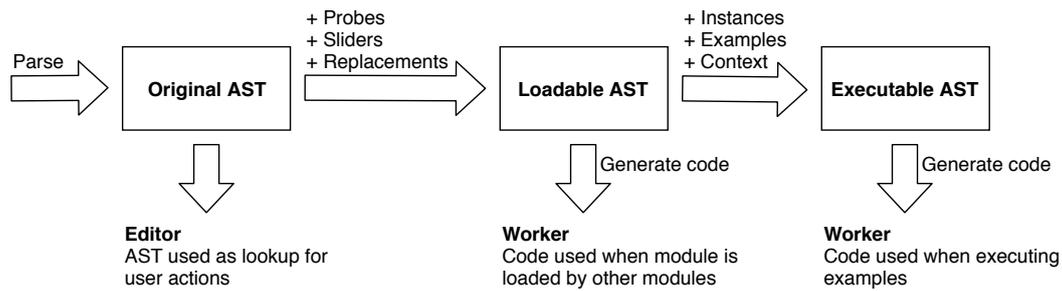

**Figure 13** The AST is transformed in several steps, and different versions are required for different tasks

## 4.1 Source Code Transformation

The first step of the evaluation process is to *parse the source code* in order to obtain the AST, which is based on the ESTree specification [11]. As the editor is text-based, the AST has to be completely regenerated after every change. While an incremental parser could increase performance [8], there are currently no widely-used incremental parsers for JavaScript. Once the AST has been generated, unique ids are assigned to all its nodes before copying the AST. This allows us to maintain different versions of the AST simultaneously (see figure 13) and still find corresponding nodes.

Before annotations are applied to the AST, some *basic transformations* are performed. First, all optionally block-scoped statements are wrapped in blocks. This ensures that the AST will stay valid as later transformations insert new statements as siblings. Second, import statements with relative paths are rewritten, as the code will be executed in a workspace and not at the file's location. Additionally, import statements referring to modules that are currently opened in other editors are rewritten to refer to the transformed versions of these modules. Third, a statement that cancels execution if the execution time exceeds a configurable value is inserted at the beginning of every block. As examples will be executed on the same thread as the editor, we want to limit the execution time to keep the editor responsive.

Finally, during the *application of annotations*, different transformations are applied to the AST depending on their type.

For probes, the values of identifiers at certain points have to be captured. As the value may change as a result of the probed statement, two statements are inserted for every probe if possible: One to capture the identifier's value before the probed statement, and one after the statement.

As sliders are concerned with the control flow, they require counters to be inserted into iterate control structures (which includes functions). These counters keep track of how often the body was entered. As the current iteration counter is also passed to probes, values can be associated with certain iterations.

---

[2] The source code of the implementation as described here is available at https://doi.org/10.5281/zenodo.2541933.

**Babylonian-style Programming**

Replacements are implemented by simply replacing the AST node they refer to with the provided alternative.

For every instance template, a factory function is generated and appended to the module. The provided constructor parameters as well as the name of the class the annotation is attached to are used to call the class's constructor. Custom instance templates are already defined as code, which is directly wrapped in a factory function.

For every example, one try-catch-block is appended to the end of the document. This block first executes the example's prescript before invoking its function and afterwards executing the postscript. All example invocations are wrapped in try-catch blocks because the occurrence of an error in one example should not prevent the execution of later examples.

## 4.2 Executing Examples

In order to collect data about the behavior of code, the worker contains a *value tracer*. This tracer is used during the execution of examples in order to log information, such as the value of an identifier at a certain point. After an example execution, the tracer provides the captured values to annotations. When an object is captured by the tracer, the tracer also logs its type and identity before copying it.

The result of all AST transformations is an executable version of the module. This *modified module is executed* within a Lively workspace. Since examples are expressed as try-catch-blocks at the end of the module's source code, they are executed as soon as the module is loaded. The rewriting of import statements (see section 4.1) ensures that if other modules are opened in the editor, their modified versions will be loaded. This means that probes will also capture values for examples that originated in another module.

Once all enabled examples have been executed, the worker informs all editors that new data is available. The editors then consult the value tracer to *update their annotations* and show new values to developers.

## 5 Programming Experience of Babylonian-style Programming

In order to illustrate the programming experience resulting from the presented design, we present different use-cases inspired by common topics in previous works. However, as most existing solutions do not aim at complex applications, we also introduce a more complex scenario in order to illustrate the additional possibilities provided by the presented design. This chapter also provides a preliminary analysis of the response time characteristics of the presented implementation.

### 5.1 Programming Experience: Binary Search, Canvas, Code Editor

We use the presented editor with three prototypical use-cases. The first one focuses on the implementation of algorithms, the second one on working with graphical output, and the third one on complex applications. The first two use-cases, binary search and canvas drawing, allow for a comparison of our approach to existing systems. The third use-case illustrates the programming experience of our approach when working with applications implemented in layers each containing several modules.

#### 5.1.1 Binary Search

As a fundamental algorithm, binary search is a common topic in programming education. This makes it an attractive example for LPEs with explicit examples aiming to make code more understandable and learnable. It is therefore used as a demonstration in several other works (see figure 14 for screenshots of different systems).

**Other Designs**   The general code editor presented in the Inventing on Principle demonstration presents an IDE-integrated two-pane solution [35]. As this design does not support probes, it shows the intermediate values of variables. If there are multiple values for the same identifier they are organized in columns, making then easily comparable for small use-cases. The design does not support examples, so the entire application is executed. The Live Literals system proposes a language-integrated single-pane design [33]. As a single-pane design, examples and intermediate values are more tightly integrated into the source code. This makes it easier to read them in the flow of the code. The lack of special highlighting does, however, also make it hard to distinguish between the abstract code and the concrete example. As multiple values for iterations are not organized in uniform columns, it is hard to find corresponding values.

**Comparison**   When we apply the presented design to the same problem, we find some significant differences to other systems. For one, the IDE-integrated, single-pane design displays information in the context of the source code, but also ensures that there is a clear visual distinction between code and examples. It is also possible to define more than one example for a function, allowing programmers to compare the behavior for different examples. While multiple values in probes are not organized in uniform columns, sliders can be used to focus on a single iteration and find corresponding values. As sliders can also be attached to functions, the presented design can also be used for a recursive implementations of the same algorithm. We can also examine a use-case for instance templates: instead of defining the array to be searched explicitly within the code, it was defined as an instance template with a descriptive name. This also makes it reusable for multiple examples.

#### 5.1.2 Canvas

LPEs with explicit examples can be used not only to gain insight into the behavior of abstract business logic, but also to get more immediate feedback when working in visual domains. We therefore examine the scenario of an application drawing a scene onto a canvas element, a common demonstration for LPEs [36, 35]. Screenshots of different systems are provided in figure 15.

It should be noted that this scenario focusses on producing a static image. There are also works aimed at providing live results for animated images [35, 23, 22], especially in the context of reversible debugging [5].

**Babylonian-style Programming**

```
function binarySearch (key, array) {             key   = 'g'
                                                 array = ['a','b','c','d','e','f']
    var low = 0;                                 low  = 0
    var high = array.length - 1;                 high = 5

    while (low <= high) {                        low   = 0  | 3  | 5
                                                 high  = 5  | 5  | 5
        var mid = floor((low + high)/2);         mid   = 2  | 4  | 5
        var value = array[mid];                  value = 'c'| 'e'| 'f'

        if (value < key) {
            low = mid + 1;                       low = 3  | 5  | 6
        }
        else if (value > key) {
            high = mid - 1;
        }
        else {
            return mid;
        }
    }

    return -1;                                   return -1
}
```

**(a)** An implementation of binary search in *Inventing on Principle* [35]

```
function binarySearch(key, array) {
    run({key: 'g', array: ['a', 'b', 'c', 'd', 'e', 'f']});

    var low = 0;
    var high = array.length - 1;

    while (low <= high) {
        p(low, [0,3,5]);
        p(high, [5,5,5]);

        var mid = Math.floor((low + high)/2);
        var value = array[mid];

        p(mid, [2,4,5]);
        p(value, ["c","e","f"]);

        if (value < key)
            low = mid + 1;
        else if (value > key)
            high = mid - 1;
        else
            return mid;
    }
    return -1;
```

**(b)** An implementation of binary search in *Live Literals* [33]

```
function binarySearch(key, array) {
    ✗ ⊙ ⇄ ⋯  Found   this: null ⇅%  key: "e" ⇅%  array: "a" to "f" ⇅%
    ✗ ⊙ ⇄ ⋯  Not Found  this: null ⇅%  key: "g" ⇅%  array: "a" to "f" ⇅%
function search(low, high) {
    ✗  Not Found  ——●——  3 of 4
        ✗ Q low:  Not Found  5
            ✗ Q high:  Not Found  5
    if(low > high) {
        return -1;
        ✗ Q return:  Not Found  /
    }
    var mid = Math.floor((low + high) / 2);
    var value = array[mid];
        ✗ Q value:  Not Found  f

    if(value < key)
        return search(mid + 1, high);
        ✗ Q return:  Not Found  -1
    else if (value > key)
        return search(low, mid - 1);
    else
        return mid;
    }
    return search(0, array.length - 1);
    ✗ Q return:  Not Found  -1
}
```

**(c)** A recursive implementation of binary search in the presented design

■ **Figure 14** A comparison of binary search in different systems

**David Rauch, Patrick Rein, Stefan Ramson, Jens Lincke, and Robert Hirschfeld**

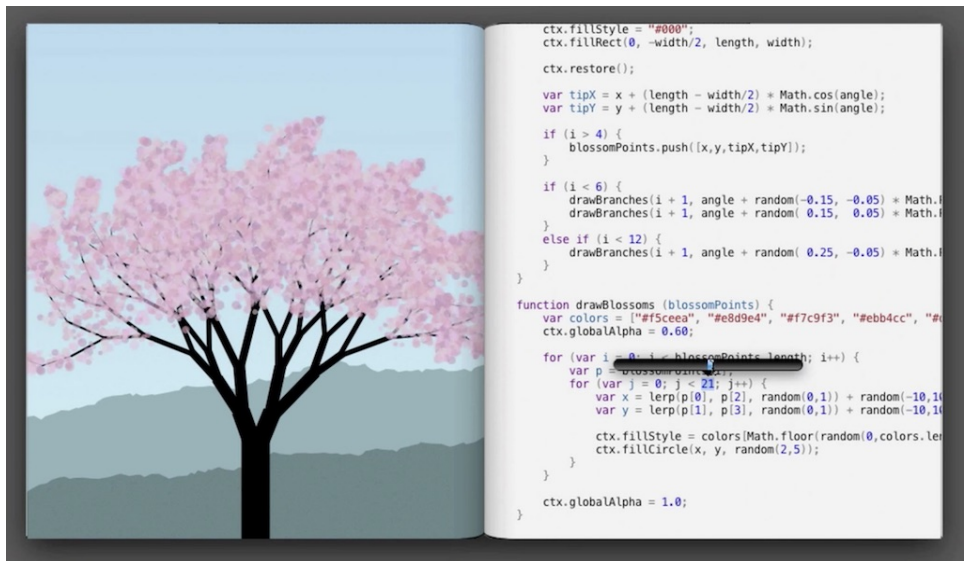

**(a)** The tree scene demo from *Inventing on Principle* [35]

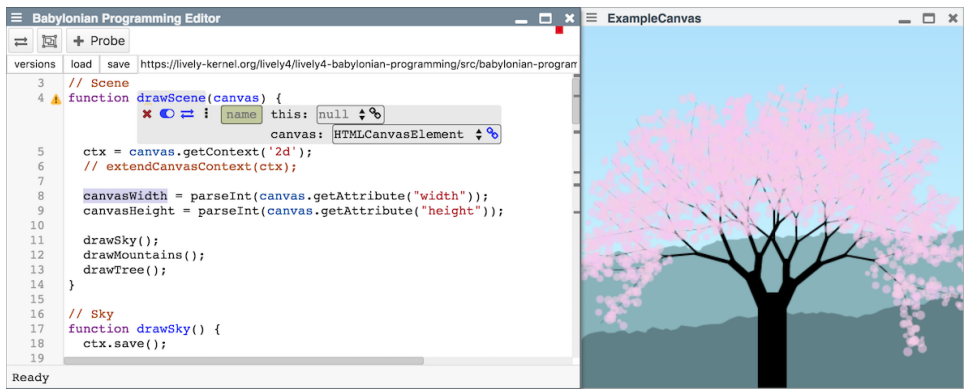

**(b)** The tree scene demo in the presented design

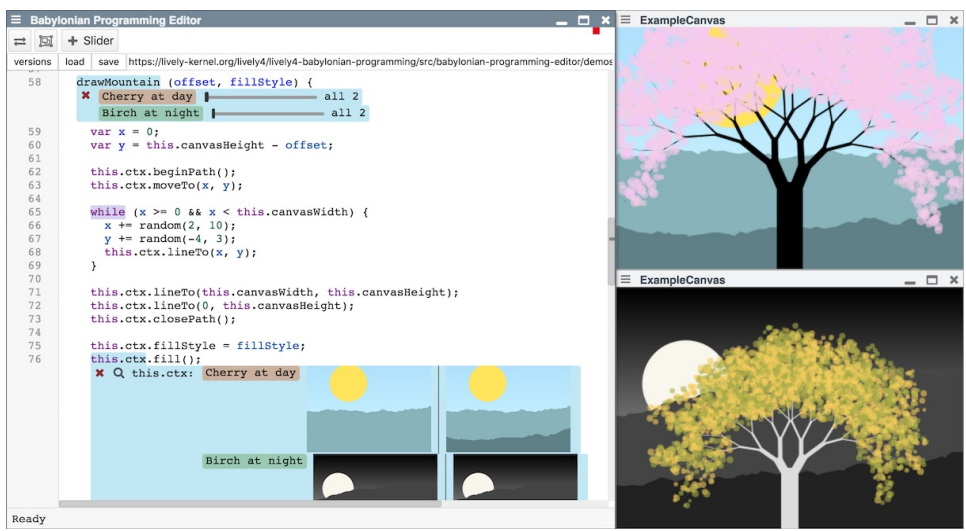

**(c)** A modular tree scene implementation in the presented design

**Figure 15** A comparison of the tree scene in different systems

**Babylonian-style Programming**

**Other Designs**   One demo in Inventing on Principle [35] presents an editor where the left side shows a live image while the right side contains a text editor. This design allows programmers to modify numerical values by dragging and to select areas in the image to jump to the code that produced them.

The design is, however, highly specialized for its domain and does not support explicit examples. For example, code presented in the editor produces one concrete image and assumes that the canvas already exists. More complex applications, which for example directly accept user input, would most likely require a more abstract implementation. This would be impossible to produce in this design, as there is no way to provide the required parameters.

**Comparison**   When the presented design is applied to the same scenario, programmers can define an example for the `drawScene` function. A link can then be used to connect the `canvas` parameter to any canvas element within the LPE. This canvas will then show a live version of the image produced by the example.

On one hand, the general-purpose nature of the tool design means that it does not provide some of the features that designs specialized for a domain offer. For example, it is currently not possible to select an area of the image to jump to the code that produced it. Even if our design supported relating program output to executed program statements, we would still have to provide a domain-specific view supporting this interaction for graphical output. On the other hand, this approach also means that the design makes no assumptions about the main entry point of the application. Programmers can define examples for functions other than `drawScene`, connect them to other canvases, and examine different parts of the application. Additionally, as probes also capture the types of objects, the canvas element can be probed at different points in order to get a preview of its contents within the editor.

The presented design can also be applied to a more modular implementation of the same scenario. In this implementation, the provided functionality is encapsulated in a class which can draw a variety of tree scenes.

Using instance templates, different instantiations of this class can be described. Examples can then be used to draw different scenes on different canvas elements simultaneously, allowing programmers to compare the behavior of code depending on its use within the system.

### 5.1.3 A Complete Editor

As a final evaluation, we aim to apply the presented design to its own implementation, therefore creating a self-documenting system.

To explore possible features when documenting a complex application, we will examine the function `generateLocationMap`, which, given an AST, generates a lookup table for AST nodes by location. This function has several properties that would make it hard to examine using existing solutions: It *recursively traverses* the AST using a visitor pattern, invokes functions defined in *other modules*, and *modifies properties* of *complex objects*.

Three examples are shown in figure 16, one of which purposefully causes an error. Since we can not define a whole AST inline, we use custom instance templates which create an ASTs by parsing short code snippets.

```
function generateLocationMap(ast) {
    ✗ ◉ ⇄ ⋯   Simple    this: null ⇅% ast: Simple AST ⇅%
    ✗ ◉ ⇄ ⋯   Fibonacci this: null ⇅% ast: Fibonacci AST ⇅%
    ✗ ◉ ⇄ ⋯ ⚠ Not an AST this: null ⇅% ast: <div> ⇅%
```

**Figure 16** Three examples defined for generateLocationMap

We can immediately see that the invalid example produces an error, and hovering over the icon allows us to read the exact error message. This example documents that this functions throws an error for invalid input rather than, for example, handling the error and returning a special value.

We continue to examine the function and want to investigate how the AST is traversed. We know that the traversing function is called repeatedly since the AST is traversed using a visitor pattern. Thus, we attach a slider to the function within the visitor object and also apply probes to identifiers of interest. The result of this is shown in in figure 17.

```
traverse(ast, {
  enter(path) {
    ✗  Simple    ——●————   2 of 5
       Fibonacci ——————●—  9 of 23
    let location = path.node.loc;
        ✗ 🔍 path.node: Simple   😺 {type: VariableDeclaration, start: 0, end: 11,
                       Fibonacci 🐱 {type: NumericLiteral, start: 29, end: 30,
    if(!location) {
      return;
    }
```

**Figure 17** Attaching a slider to a function within an visitor object to examine AST traversal

Scrolling through the activations and examining the `type` property within the probe, we learn that nodes are traversed in pre-order. We could also attach a probe to the `AST` identifier and open the captured value in the object editor to examine the AST in tree form. We also see that there was no node without a location, as the conditional code was never executed.

To see how the location map is generated, we apply a probe to it while it is modified, as shown in figure 18. Scrolling through iterations, we can see how keys are added and existing values are overwritten. As this object is rather complex, the probe only presents us with a flat representation. We can, however, open any object in the object inspector for closer inspection.

```
ast._locationMap[LocationConverter.astToKey(location)] = path;
✗ 🔍 ast._locationMap: Simple   😺 {toJSON: [object Object], 1,0,1,11: [object Object] [object Ob
                       Fibonacci 🐱 {toJSON: [object Object], 1,0,5,1: [object Object], 2,0,5,1
```

**Figure 18** Attaching a probe to a complex object while it is modified

**Babylonian-style Programming**

Additionally, the map's keys are not very descriptive. As they are generated by the class `LocationConverter`, we open it in a second editor. Since `astToKey` is the only function that is called by our enabled examples, all other functions within `LocationConverter` are faded out. This helps to quickly find code that is relevant to the task at hand.

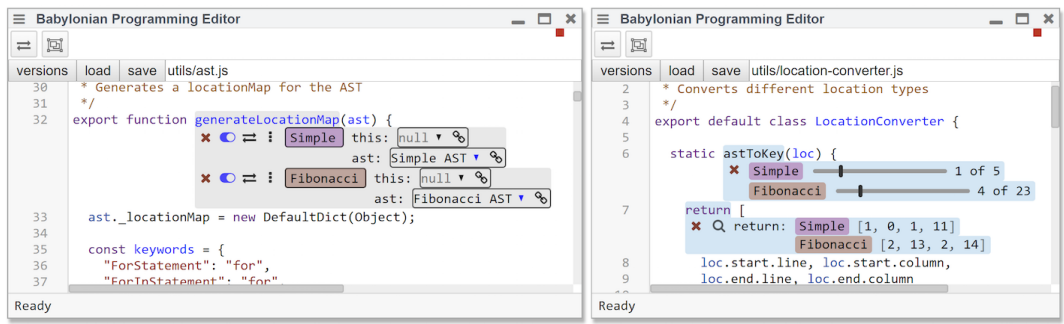

■ **Figure 19** A probe capturing values for examples from different modules

The relevant function already has an example defined, as shown in figure 19. The probe can, however, also show values for our examples from another module, as shown in figure 20. This allows us to follow our example from our initial module to the module of `astToKey` and quickly see how the keys are generated in our scenario.

■ **Figure 20** Probes in one editor capture values for enabled examples defined in another editor.

This scenario shows how the presented design can be used to inspect complex, changing values over time in order to gain insight into abstract business logic with complex flow control structures.

### 5.2 Preliminary Response-time Evaluation

The live programming experience partially depends on an impression of immediacy of feedback. This impression of immediacy in turn depends on the time between a change to the source code or example and an update in the feedback mechanism [14, 29].

In order to explore the potential impact of our design, we determined the duration of the feedback cycle programmers would experience in different scenarios (see appendix A): An empty editor as a baseline, a simple algorithm with three enabled

9:26

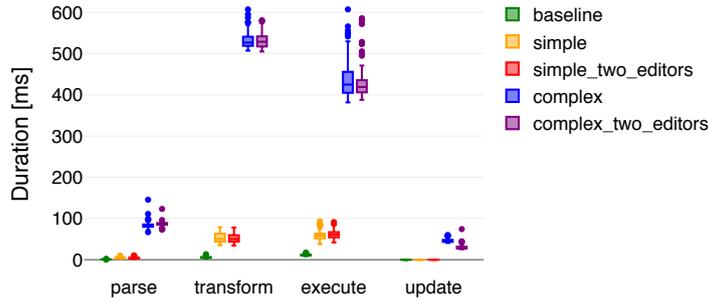

**Figure 21** A box plot showing the whole range of durations for different steps of processing

**Table 1** The median duration and standard deviation for different steps of processing provided in ms

|  | Adaptation | | Emergence | |
| --- | --- | --- | --- | --- |
|  | parse | transform | execute | update |
| baseline | 0.50 ± 0.15 | 5.20 ± 1.57 | 11.30 ± 1.61 | 0.00 ± 0.04 |
| simple | 3.80 ± 2.09 | 50.50 ± 11.81 | 58.55 ± 10.22 | 0.10 ± 0.06 |
| simple_two_e. | 3.80 ± 2.13 | 50.15 ± 11.63 | 60.80 ± 10.61 | 0.10 ± 0.06 |
| complex | 83.15 ± 8.60 | 527.05 ± 19.90 | 425.10 ± 46.69 | 45.95 ± 4.26 |
| complex_two_e. | 86.95 ± 5.40 | 529.05 ± 18.23 | 419.20 ± 40.82 | 29.25 ± 5.86 |

examples and a large module with eight enabled examples. We also evaluated the effect of opening a second editor that is affected by these examples.

The measurements are split into four steps. The adaptation phase between source code and the adapted executable version is split into the "parse" and the "transform" steps [29]. This phase can be performed on a separate thread. The "execute" and "update" steps represent the "emergence" phase, as they start with the executable module and end with the updated results being presented to users. This phase has to be performed on the main thread, meaning that the editor is unresponsive during this phase.

**Method** All measurements were performed using the Performance API, which provides wall-clock time measurements accurate to 5 microseconds [27]. We measured wall-clock time as we were interested in the delay programmers would experience when waiting for feedback. We repeated the measurements 100 times per scenario. The browser tab was kept open and in the foreground. In order to simulate a programmer working on an application, a change-event was fired every five seconds, triggering a re-evaluation of all enabled examples. In between scenarios, the system was reset by clearing all browser caches and restarting the browser. There were no other applications running on the same system except Operating System (OS)-related services (see appendix C for details on the benchmark system).

**Results** First of all, several results show a comparatively high standard deviation. As we are measuring wall-clock time in a web browser with a default configuration, the measured durations include time spent on other activities which might only be indirectly related to our scenarios, for example layouting or just-in-time compilation. This is intentional, as we aim to measure the actually experienced response times including any activities of the browser which might be influenced by our implementation.

Examining the results presented in table 1, we find that the transformation of the AST and execution of examples require the most time, while parsing is generally fast, and updating the UI is only relevant for the complex scenario. We find that for a low number of enabled examples, the overall duration is slightly above the threshold of 100 ms for immediacy, but far below the critical 1 s threshold [14]. However, as more examples are enabled, the overall duration exceeds the 1 s threshold.

This preliminary evaluation showed that the implementation is limited by the possibilities of web components. The fundamental limitation of executing examples on the same thread as the UI presented challenges when working with multiple enabled examples on large projects. These limitation are, however, inherent to the implementation and not the design itself.

**Threats to Validity** We identified several threats to the validity of the results as well as to the conclusions. First, as only three scenarios were examined, the results should only be seen as preliminary. Second, measurements may have been influenced by the warm-up phase of the JavaScript just-in-time compiler. Repeatedly executing the same code may not represent real-world usage. It may allow for more optimizations, and therefore lower the average duration. Additionally, more user studies are necessary to draw definitive conclusions about the usability of the implementation. For example, examples are only evaluated once programmers stop typing for some time, it could be possible that the response times are acceptable for quickly typing programmers but not for programmers who type slowly.

### 5.3 Comparison of Features for Working with Explicit Examples

Initially, we identified a set of requirements based on features and limitations of existing systems (see section 2.2). In order to illustrate how our proposed design fits into the existing spectrum and what future challenges remain, we compare our approach to these requirements and the features of the other existing systems (see table 2).

### 5.4 Discussion: Upcoming Challenges

The comparison in table 2 shows that our approach integrates features of existing environments and adapts them for working with larger code bases. At the same time several requirements remain unfulfilled.

While the presented design allows programmers to enable multiple examples simultaneously and examine the resulting behavior in other modules, the resulting overall trace is not always obvious. For example, if code within a low-level module is executed as a result of two examples in a high-level module, probes within the low-level module will show values for both examples. They will, however, not indicate

**Table 2** Requirements for LPEs with explicit examples for complex applications and the features satisfying them in different systems.

| requirements | Proposed Tool Design | Example Centric Programming | Live Literals | Shiranui | Inventing on Principle: Canvas | Inventing on Principle: General Code | Learnable Programming | Seymour | Light Table |
|---|---|---|---|---|---|---|---|---|---|
| automatic feedback granularity | probe on any expression in program code | result of every statements | probes on provided expressions | probes on provided expressions | program output | result of every statement | result of every statement | result of every statement | probe on any expression in program code |
| state over time | all states of probe inline, explicit state transitions, sliders to navigate to points in time | trace of all statements | all states of probe inline | all states of probe inline | - | all results of statements in second pane at once | all results of statements in second pane at once, navigation by horizontal scrolling | all results of statements in second pane at once, macro view to navigate to points in time | last result of watch expression |
| state over modules | all probes throughout the system show results | navigating the full trace | not explicitly covered, potentially possible | not explicitly covered, potentially possible | not applicable | not explicitly covered | not explicitly covered, potentially possible | not explicitly covered, potentially possible | all watches across files show results from example evaluation |
| arbitrary objects | object identity through emoticons, complex state in probe widget, object inspector on demand | not explicitly covered | state printed inline | not explicitly covered | not applicable | not explicitly covered | not explicitly covered | object identity through emoticons, object state not explicitly covered | complex object state in watch widget |
| domain-specific feedback | none | none | none | none | program output as only feedback | none | provided for graphical objects | none | provided for some objects such as graphs |
| multiple examples | multiple explicit examples, named examples, individually enabled | implicit examples, selection through trace | explicit single example, multiple tests possible | explicit examples for testing, selecting one for probes | implicit single example | explicit single example | implicit single example | implicit multiple examples, selection through macro view | implicit multiple examples, one active for watches |
| reusing example parts | through named instance templates, links to objects | through variables in the program | through variables in the program | not explicitly covered | not possible | not explicitly covered | not possible | through variables in the program | through variables in the program |
| behavioral highlighting (trace) | executed statements per example (trace) | executed statements per example (trace) | none | statements relevant to some state (slice) | from program output to code | none | none | executed statements per example (trace) | last execution of watch expression |
| specifying context | automatically executed pre- and postscripts, replacements, executions partially isolated, overriding of explicit "holes" in code | explicit setup code before implicit example | no mechanism, isolation not explicitly covered | no mechanism, isolated executions | not applicable | no mechanism | no mechanism | explicit setup code before implicit example | explicit setup code before implicit example |
| keeping track of assumptions | shows erroneous examples | locations of exceptions, assertions in trace view | asserting the result of example | asserting the result of example | none | none | none | not explicitly covered | details on exceptions inline |
| navigating the trace | sliders for iterations and activations | full trace view for navigation | none | through slices by focusing on a result | none | none | none | macro view for trace navigation | none |

9:29

whether both examples arrived at this point via the same execution path. Future work may therefore examine ways to display and compare execution paths. One possible way to present the global execution path can be found in Seymour's macro view [15]. Another potential way could be the inline stack navigation of YinYang [24].

Another future challenge for our tool design are domain-specific visualizations which do not break the close integration of examples, the presentation of dynamic information, and source code. Source code editor augmentation approaches such as in-situ visualization show that this is possible even for detailed data visualizations and that users can benefit from it [12, 34]. Projectional editors might be another approach to enabling to integrate visualizations and source code [37, 31].

In addition, the integration of mined real-world traces such as proposed by the Tralfamadore system might be beneficial [19]. Programmers could then first investigate real-world traces and then explore implementation alternatives and see the effects on the program behavior directly.

A more general challenge are examples in non-terminating systems, such as the game loop of a game [24]. Programmers might want to determine how certain behavior in the game relates to internal state. Our tool design can not be applied to such scenarios. Future work may examine whether techniques for non-terminating examples from systems with implicit examples are also suitable for systems with explicit examples [5, 9, 22, 35].

In order to encourage the adoption of example-based software development, the integration of unit tests and examples should be further examined. Unit tests and examples in live programming can both be seen as curated examples and thus tests could be considered as a source of examples, and examples could be used as templates for unit tests.

Beyond future design challenges, future work might also investigate the differences and commonalities between requirement spaces of approaches related to live programming with examples. For example, while the design principles for learnable programming aim for accessibility of programming systems, they represent similar requirements for tool designs [35]. One example for this is the principle "Follow the Flow" which is similar to our requirement "Navigating the Trace".

## 6 Conclusion

The integration of live examples into source code editors may support the comprehension, exploration and development of applications by closely connecting the static source code of an application with its dynamic behavior [20]. In an exploration of existing solutions, we identified a number of requirements to work with complex applications with extensive code bases. We have therefore proposed a new design targeting larger software projects. The presented design integrates behavioral information into code and supports various ways to either create an appropriate context before examples are evaluated or to use existing parts of the LPE as context for example execution.

An evaluation of the proposed design showed that it can provide functionality similar to existing designs without being limited to small applications or specific domains. Further scenarios demonstrated the additional possibilities provided by the new design for complex applications, such as the possibility to examine the evaluation of examples across multiple interacting modules. A performance evaluation showed that the current implementation is limited by some aspects of the underlying architecture. As this bottleneck is inherent to the implementation's environment and not the design itself, future implementations on other platforms may provide significant performance improvements.

**Acknowledgements**   We gratefully acknowledge the financial support of the Research School for Service-oriented Systems Engineering of the Hasso Plattner Institute and the Hasso Plattner Design Thinking Research Program. Sincere thanks also goes to the anonymous reviewers for their detailed feedback.

## A  Benchmark Scenarios

### A.1  Simple Benchmark Scenario

Listing 1 and listing 2 show the code used as the simple benchmark scenario. A full version of the code including all example comments is available as part of the published source code.

**Listing 1**  The content of simple.js with shortened example comments

```javascript
1  import Person from "./person.js";
2
3  function /* 3 examples */binarySearch(array, element, compare) {
4      var low = 0;
5      var high = array.length - 1;
6      /* slider */while (low <= high) {
7          var mid = (low + high) >> 1;
8          var /* probe */compareResult = compare(element, array[mid]);
9          if (compareResult > 0) {
10             low = mid + 1;
11         } else if(compareResult < 0) {
12             high = mid - 1;
```

```
13        } else {
14            /* probe */return mid;
15        }
16    }
17    /* probe */return -1;
18 }
```

■ **Listing 2**  The content of person.js with shortened example comments

```
1 export default class Person {
2   constructor(name) {
3     this.name = /* probe */name;
4   }
5
6   sayHi() {
7     console.log(`Hi, Im \$\{this.name\}`);
8   }
9 }
```

### A.2  Complex Benchmark Scenario

Listing 3 and listing 4 show excerpts of the code used as the complex benchmark scenario. A full version of the code including all example comments is available as part of the published source code.

■ **Listing 3**  The partial content of ast.js with shortened example comments

```
1  /* Imports... */
2
3  export function /* 3 examples */deepCopy(obj) {
4    try {
5      /* probe */return JSON.parse(JSON.stringify(obj));
6    } catch(e) {
7      console.warn("Could not deeply clone object", obj);
8      /* probe */return Object.assign({}, obj);
9    }
10 }
11
12 export function /* 2 examples */generateLocationMap(ast) {
13   ast._locationMap = new DefaultDict(Object);
14   const keywords = { /* List of keywords */ };
15   traverse(ast, {
16     /* slider */enter(path) {
17       let location = path.node.loc;
18       if(!location) {
19         return;
20       }
21       // Some Nodes are only associated with their keywords
22       const keyword = keywords[/* probe */path.type];
23       if(keyword) {
```

```
24            location.end.line = location.start.line;
25            location.end.column = location.start.column + /* probe */keyword.length;
26          }
27          ast._locationMap[LocationConverter.astToKey(location)] = path;
28        }
29      });
30    }
31
32    export function /* 2 examples */canBeProbe(path) {
33      if(!path) {
34        return false;
35      }
36      const /* probe */isTrackableIdentifier = /* Expression… */;
37      const isTrackableParameter = /* Expression… */;
38      const /* probe */isTrackableMemberExpression = path.isMemberExpression();
39      const /* probe */isTrackableReturnStatement = path.isReturnStatement();
40      /* probe */return isTrackableIdentifier || isTrackableParameter
41                    || isTrackableMemberExpression || isTrackableReturnStatement;
42    }
43
44    export function /* example */canBeSlider(path) {
45      if(!path) {
46        return false;
47      }
48      const isTrackableIdentifier = path.isIdentifier()
49                              && path.parentPath === path.getFunctionParent();
50      const isTrackableLoop = path.isLoop();
51      return /* probe */isTrackableIdentifier || /* probe */isTrackableLoop;
52    }
53
54    /* 550 more lines… */
```

■ **Listing 4** The partial content of location-converter.js with shortened example comments

```
1   export default class LocationConverter {
2
3     static /* slider, example */astToKey(loc) {
4       /* probe */return [
5         loc.start.line, loc.start.column,
6         loc.end.line, loc.end.column
7       ];
8     }
9
10    /* 50 more lines… */
11
12  }
```

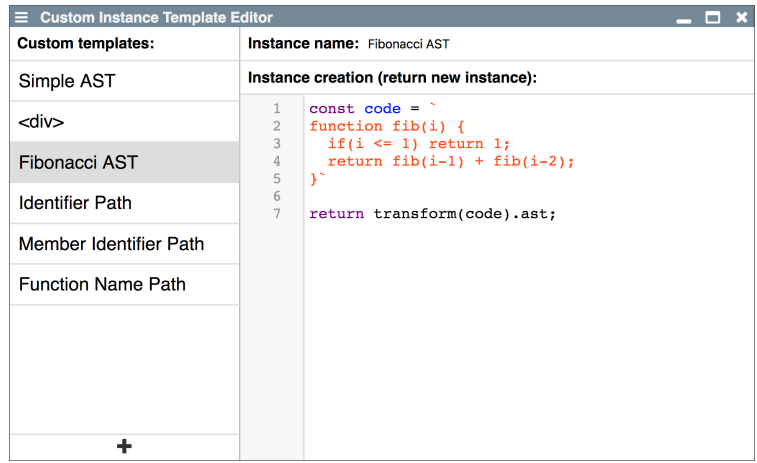

**Figure 22** Several instance templates shown in the custom instance template editor

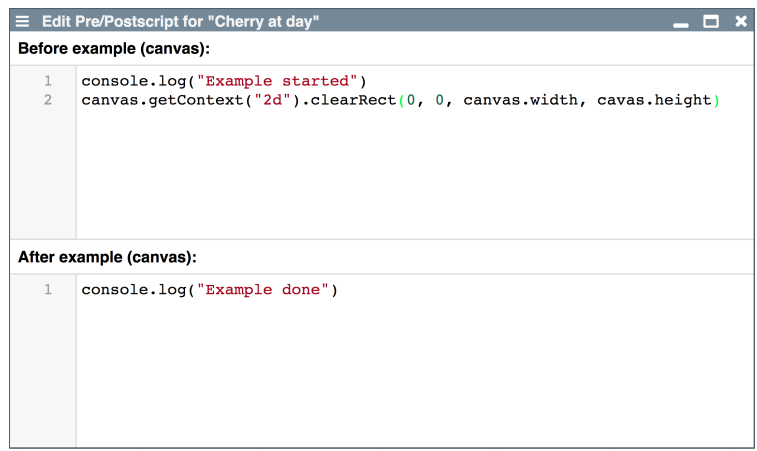

**Figure 23** The Pre/Postscript editor for an example that requires a canvas to be cleared before execution

## B  Additional UI Screenshots

This section provides screenshots of additional UI components. The custom instance template editor is visible in figure 22, while figure 23 shows the Pre- and Postscript editor and figure 24 shows the object inspector.

## C  Benchmark System Specification

The benchmarks described in section 5.2 were conducted using the following system:
- Browser: Google Chrome, Version 67.0.3396.99 (64-bit)
- OS: macOS High Sierra, Version 10.13.5 (Build 17F77)
- CPU: Intel® Core™ i7-7700HQ, 2.80 GHz, 8 logical cores
- Memory: 16 GB LPDDR3, 2133 MHz

**Babylonian-style Programming**

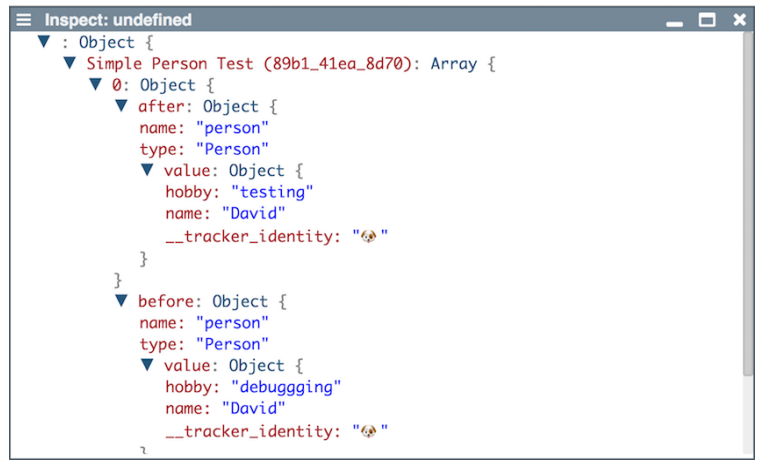

■ **Figure 24** The object inspector showing the data captured by a probe

- GPU: Radeon Pro 555, 2048 MB GDDR5 Memory
- Connected to power, discrete GPU activated

9:38

## About the authors

**David Rauch** is a graduate student at the Software Architecture Group of the Hasso Plattner Institute at the University of Potsdam. His research interests include programming tools and live programming in particular. Contact David at mail@davidrauch.at.

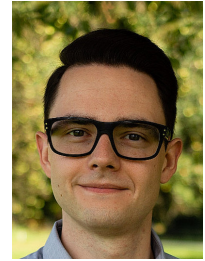

**Patrick Rein** is a PhD student in the Software Architecture Group of the Hasso Plattner Institute at the University of Potsdam. His research interests include live and exploratory programming systems as well as personal information management systems. Contact Patrick at patrick.rein@hpi.uni-potsdam.de.

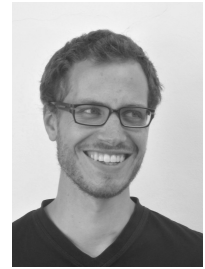

**Stefan Ramson** is a doctoral researcher at the Software Architecture Group. His research interests include programming language design and natural programming. Contact Stefan at stefan.ramson@hpi.uni-potsdam.de.

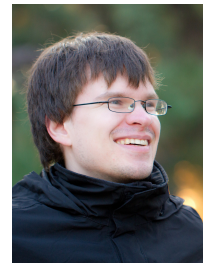

**Jens Lincke** is a member of the Hasso Plattner Institute's Software Architecture Group. His research interests include live and exploratory programming. Lincke received a PhD in IT-Systems Engineering from the Hasso Plattner Institute at the University of Potsdam. Contact Jens at jens.lincke@hpi.uni-potsdam.de.

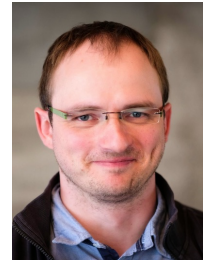

**Robert Hirschfeld** leads the Software Architecture Group of the Hasso Plattner Institute at the University of Potsdam. His research interests include dynamic programming languages, development tools, and runtime environments to make live, exploratory programming more approachable. Hirschfeld received a PhD in computer science from Technische Universität Ilmenau. Contact Robert at robert.hirschfeld@hpi.uni-potsdam.de.

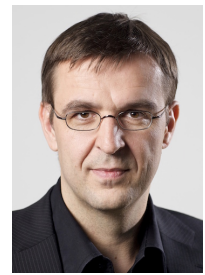